# Méthode de calcul du rayonnement acoustique de structures complexes

**Marianne Viallet\*,\*\* — Gérald Poumérol\* — Olivier Dessombz \*\* — Louis Jézéquel \*\***

*\* PSA Peugeot Citroën, Direction des Plateformes Techniques et des Achats*
*Centre Technique de la Garenne Colombes — Case courrier LG098*
*18, rue des Fauvelles, 92256 La Garenne Colombes cedex*
*marianne.viallet@mpsa.com*

*\*\* Laboratoire de Tribologie et Dynamique des Systèmes, École Centrale de Lyon*
*CNRS UMR 5513*
*36, avenue Guy de Collongues, 69134 Ecully cedex*

RÉSUMÉ. *Dans l'industrie automobile, prédire le bruit rayonné est une étape importante de la conception. Pour résoudre les problèmes acoustiques, il existe principalement deux familles de méthodes : les Méthodes Éléments Finis (FEM) et les Méthodes Éléments de Frontières (BEM). Pour calculer le rayonnement acoustique en champ libre, on utilise généralement plutôt les éléments de frontières. Néanmoins ces méthodes peuvent induire des singularités, et sont par conséquent moins faciles à utiliser que les éléments finis qui eux sont plutôt adaptés à l'étude des milieux bornés. La méthode décrite dans cet article, la SDM, permet de tirer avantage de ces deux méthodes en utilisant chacune là où elle est la plus performante. Une nouvelle méthode fondée sur éléments finis est également présentée et permet de remplacer avantageusement les éléments de frontière pour traiter le problème extérieur. L'efficacité de la SDM couplée à cette nouvelle méthode est discutée.*

ABSTRACT. *In the automotive industry, predicting noise during design cycle is a necessary step. Well-known methods exist to answer this issue in low frequency domain. Among these, Finite Element Methods, adapted to closed domains, are quite easy to implement whereas Boundary Element Methods are more adapted to infinite domains, but may induce singularity problems. In this article, the described method, the SDM, allows to use both methods in their best application domain. A new method is also presented to solve the SDM exterior problem. Instead of using Boundary Element Methods, an original use of Finite Elements is made. Efficiency of this new version of the Substructure Deletion Method is discussed.*

MOTS-CLÉS : *SDM, FEM, BEM, rayonnement acoustique, structures à géométries complexes, algorithme de clonage, sous structuration infinie.*

## 1. Présentation de la méthode SDM ou sous structuration soustractive

Dans l'industrie automobile, la prédiction du bruit rayonné est une étape nécessaire pendant le cycle de conception. En effet, en plus des normes les véhicules doivent satisfaire les attentes des futurs passagers. Dans ce but, et afin de réduire les coûts de retouche plus importants à la fin de la conception qu'au début, il peut être intéressant d'avoir des méthodes adaptées à l'avant projet. Nous avons ainsi proposé d'utiliser la Substructure Deletion Method, SDM ou encore appelée sous structuration soustractive. Cette méthode introduite à la fin des années 70 pour le génie civil permet de calculer la réponse dynamique d'une fondation enterrée sous chargement sismique en décomposant le problème en deux autres plus simples à résoudre (Dasgupta, 1980) (Betti *et al*, 1994). Ainsi pour résoudre ce problème, on calcule l'impédance du sol non excavé à l'emplacement de la fondation, puis celle associée au volume de sol à retirer pour placer cette dernière. En recombinant les deux matrices ainsi obtenues, on calcule l'impédance associée à la fondation et qui permet de remonter aux déplacements induits par la sollicitation sismique. Le principe est le même en acoustique. On englobe la géométrie complexe avec une surface régulière. On traite alors les deux problèmes résultants séparément (Viallet *et al*, 2006a) (Viallet *et al*, 2006b)

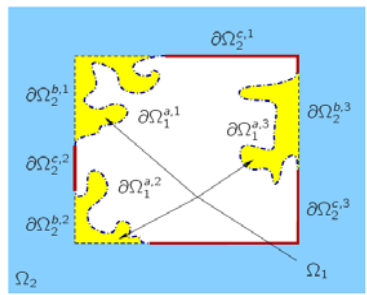

**Figure 1.** *Description de la décomposition d'un problème à géométrie complexe en deux problèmes plus simples*

## 2. Présentation de l'algorithme de clonage

Afin de remplacer l'usage des éléments de frontière pour la résolution du problème infini, nous proposons une nouvelle méthode itérative. Il s'agit d'un algorithme de clonage, aussi appelé sous structuration infinie. Le concept a été introduit pour le génie civil à la fin des années 70 (Dasgupta, 1982), (Dasgupta, 1984) et repris au début des années 90 (Benembarek, 1993) (Wolf, 2003).

La méthode consiste à diviser l'espace infini en une infinité de sous éléments finis. Ces cellules grossissent au fur et à mesure qu'on se rapproche de l'infini suivant un rapport de similitude $g$ comme cela est illustré Figure 2. Il suffit de n'étudier que la première cellule $W_1$ pour calculer l'impédance de la frontière $\partial W_1$ et par voie de conséquence résoudre le problème extérieur.

On considère le domaine $W_1$. L'équation de Helmholtz dans ce volume écrit à l'aide de la formulation des Éléments Finis nous donne l'expression suivante :

$$S^{(1)}p^{(1)} = r_0 C v_n^{(1)} = C_v^{(1)} \text{ avec } S^{(1)} = K^{(1)} - w^2 M^{(1)} \qquad [1]$$

$$\begin{bmatrix} S_{11}(w) & S_{12}(w) \\ S_{21}(w) & S_{22}(w) \end{bmatrix} \begin{Bmatrix} p_1 \\ p_2 \end{Bmatrix} = \begin{Bmatrix} C_{v\,i}^{(i)} \\ C_{v\,i+1}^{(i)} \end{Bmatrix} \qquad [2]$$

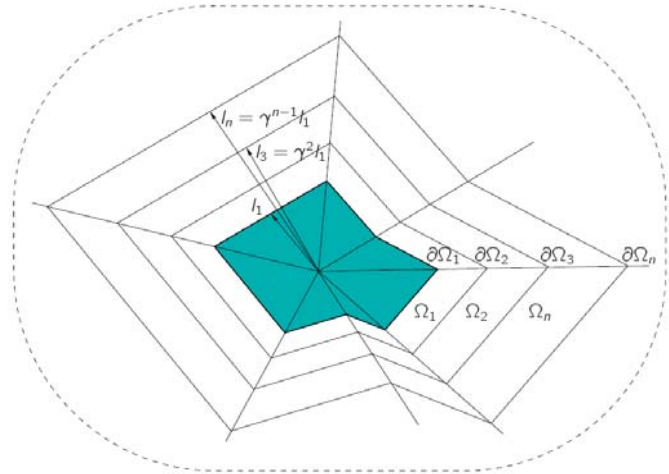

**Figure 2.** *Division de l'espace en éléments finis similaires*

On recherche la matrice d'impédance à l'infini $D^{(1)}$ telle que : $v_{n1}^{(1)} = D^{(1)} p_1$ sur $\partial W_1$. De même $D^{(2)}$ est définie comme suit : $v_{n2}^{(2)} = -v_{n2}^{(1)} = D^{(2)} p_2$. Le système [2] devient alors :

$$S_{11}(w)p_1 + S_{12}(w)p_2 = D^{(1)} p_1 \qquad [3]$$

$$S_{21}(w)p_1 + S_{22}(w)p_2 = -D^{(2)} p_2$$

Soit $K$ la dimension de l'espace, $K = 2$ en 2D, et $K = 3$ en 3D. On obtient avec des changements de variables adéquats permettant d'adimensionner les différentes grandeurs :

$$Z_{11}(w)P_1 + g^{-1}Z_{12}(w)P_2 = X^{(1)}P_1 \qquad [4]$$

$$gZ_{21}(w)P_1 + Z_{22}(w)P_2 = -g^{K-2}X^{(2)}P_2$$

Le rapport de similitude est choisi tel que $g \gg 1$ ce qui permet d'écrire $X^{(1)} \gg X^{(2)}$ et par combinaison linéaire des deux équations de [4], on parvient à l'équation suivante, quelque soient les pressions adimensionnées $P_1$ et $P_2$ :

$$X = Z_{11} - Z_{12}\left[g^{K-2}X + Z_{22}\right]^{-1}Z_{21} \qquad [5]$$

Un second changement de variables permet d'obtenir la nouvelle relation :

$$YY - QY + R = [0] \qquad [6]$$

On pose $Yy_i = l_i y_i$ où $l_i$ et $y_i$ sont respectivement les valeurs propres et les vecteurs propres de la matrice $Y$. On en déduit l'équation quadratique [7] qui est résolue en passant dans l'espace d'état.

$$l_i^2 y_i + l_i Q y_i + R y_i = \{0\} \qquad [7]$$

On obtient un jeu de 2n valeurs de $l_i$ et de 2n vecteurs $y_i$. Seuls la moitié a un sens physique, et il ne faut sélectionner que ceux qui respectent les conditions aux limites de notre problème. On remonte alors à la matrice $Y$, puis la matrice $X$ et enfin la matrice $D^{(1)}$.

## 3. Résultats

### 3.1. Validation de l'algorithme de clonage sur un cas simple

La méthode a ainsi pu être appliquée en 2D sur un cas très simple d'OPS test (One Point Source Test). Le principe est de placer une source à l'intérieur du volume délimité par la surface $\partial W_1$. Il est alors aisé de calculé la pression et le gradient de pression en tout point de l'espace grâce à la solution de Green. On utilise ensuite le gradient de pression de Green $\nabla G$ comme condition aux limites du problème et on tente de retrouver la pression de Green $G$ calculée analytiquement. L'algorithme de clonage a été testé et validé sur une structure 2D circulaire de rayon $R = 55 cm$. Seul le premier volume autour de la structure doit être maillé pour résoudre le problème de rayonnement acoustique. Plusieurs rapports de similitude $g$ ont été testés comme on peut le voir sur la Figure 3.

On constate que plus $g$ se rapproche de 1 et plus les résultats se rapprochent de la réponse analytique.

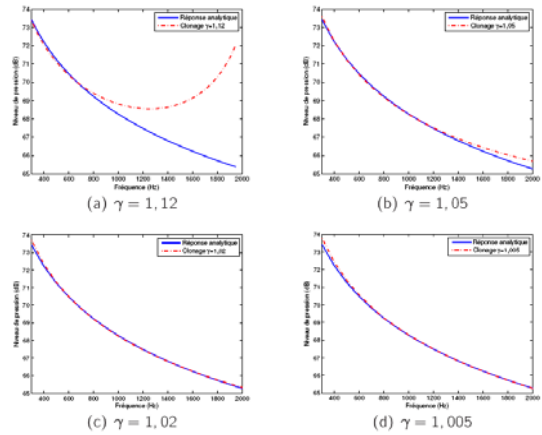

**Figure 3.** *Résultats de l'OPS Test pour différentes valeurs de gamma*

### 3.2. Utilisation de l'algorithme de clonage pour la SDM

La SDM est appliquée à un cas simple, c'est-à-dire un contour rectangulaire dont on souhaite déterminer le rayonnement. Ce contour est englobé par un contour circulaire. On applique une excitation sinusoïdale sur les nœuds du maillage fluide. Celle-ci permet de simuler déformées modales de la structure.

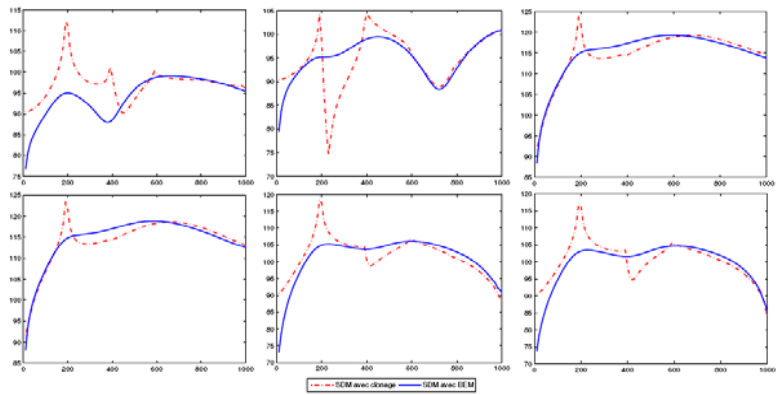

**Figure 4.** *Niveau de pression sur 6 points répartis sur la structure en dB en fonction de la fréquence*

On compare les résultats de la SDM avec les éléments de frontière pour traiter le problème extérieur et de la SDM avec l'algorithme de clonage. Globalement le comportement est assez bien respecté. Des pics apparaissent et correspondent aux modes propres de l'anneau qui viennent parasiter la réponse.

## 4. Conclusion

Nous avons présenté une méthode permettant de calculer le rayonnement acoustique de structures complexes. Celle-ci est tout à fait adaptée à la comparaison de différentes géométries ayant un même encombrement et permet de gagner dans ce cas un temps considérable. En effet, au lieu de refaire le calcul à chaque fois, il suffit alors de conserver l'impédance à l'infini du volume englobant et d'étudier le volume restant par éléments finis. Ainsi plus on compare d'architectures, plus la méthode devient intéressante.

De plus, l'utilisation de l'algorithme de clonage permet de diminuer les temps de calcul par rapport à une BEM classique.

Ces avantages font de la SDM utilisée avec l'algorithme de clonage une méthode très bien adaptée aux études en avant projet.

## 5. Bibliographie